%% file: main.tex
\documentclass[
 reprint,
 superscriptaddress,
 amsmath, amssymb,
 aps
]{revtex4-2}

\usepackage{graphicx}
\usepackage[justification=Justified,                                        singlelinecheck=off,                                              format=plain]{caption}
\usepackage[justification=justified,                                          format=plain]{subcaption}
\usepackage{dcolumn}
\usepackage{bm}
\usepackage{soul}
\usepackage[utf8]{inputenc}
\usepackage{wrapfig}
\usepackage{amsmath}
\usepackage{import}
\usepackage{amssymb}
\usepackage{float}

\begin{document}


\setlength{\belowdisplayskip}{0.25pt} 

\preprint{APS/123-QED}

\title{Using Evolutionary Algorithms to Design Antennas with Greater Sensitivity to Ultra-High Energy Neutrinos}

\input{GENETIS_institutes.tex}
\input{GENETIS_authors.tex}

\begin{abstract}
\noindent The Genetically Evolved NEutrino Telescopes for Improved Sensitivity (GENETIS) project seeks to optimize detectors in physics for science outcomes in high-dimensional parameter spaces. In this project, we designed an antenna using a genetic algorithm with a science outcome directly as the sole figure of merit. This paper presents initial results on the improvement of an antenna design for in-ice neutrino detectors using the current Askaryan Radio Array (ARA) experiment as a baseline. By optimizing for the effective volume using the evolved antenna design in ARA, we improve upon ARA's simulated sensitivity to ultra-high energy neutrinos by 22\%, despite using limited parameters in this initial investigation. Future improvements will continue to increase the computational efficiency of the genetic algorithm and the complexity and fitness of the antenna designs. This work lays the foundation for continued research and development of methods to increase the sensitivity of detectors in physics and other fields in parameter spaces of high dimensionality. 
\end{abstract}

\maketitle


\section{Introduction}
The Genetically Evolved NEutrino Telescopes for Improved Sensitivity (GENETIS) project aims to optimize the science outcomes of detector designs in high-dimensional parameter spaces to advance the field of physics. As a first application, GENETIS has produced a Genetic Algorithm (GA)~\cite{GAbasics} that evolves antenna geometries optimized for ultra-high energy (UHE) neutrino detection in a six-dimensional parameter space. GENETIS applies a heuristic optimization method for designing an antenna using a \textit{science outcome} as the sole measure of fitness. This paper presents GENETIS' initial results on the improvement of antenna designs used in UHE neutrino experiments with a limited number of parameters. 

The high-dimensional parameter spaces of detector design problems motivate using a heuristic to improve upon designs made using traditional techniques. A heuristic is a simple method for efficiently finding a high-quality solution to a given problem without evaluating all possible solutions. In particular, the design of antennas for UHE neutrino detection has explicit constraints and a high-dimensional parameter space, making it well suited for heuristic optimization. Given the immense scale of these experiments and the low flux of UHE neutrinos, each detector element must be designed to return the best science outcome for its cost. 

GENETIS chose to use GAs, among other potential computational intelligence and machine learning algorithms, for antenna design because of their effectiveness at complex optimization problems, especially when many optima could exist~\cite{Mutation}. GAs are also often more transparent than other methods, such as machine learning optimization techniques, which allows for an intuitive understanding of how the algorithm arrived at a final result. Searching the six-dimensional parameter space explored in this investigation using increments of the size necessary to find a peak fitness score would require the evaluation of more than $10^{8}$ designs. By contrast, the GA used here needed only 1550 designs to search the parameter space.

The use of GAs was initially motivated by the NASA ST-5 antenna, in which a GA designed a simple, segmented wire antenna for satellite communications~\cite{NASA_Paperclip}. Many other examples of antenna design optimization using GAs exist, including Yagi-Uda antennas~\cite{Jones97}, electrically loaded wire antennas~\cite{Boag96}, broadband cage antennas~\cite{Deng14}, planar antennas~\cite{Gulati18}, pyramid horn antennas~\cite{Deepika17}, ultra-wideband slot antennas~\cite{Xie11}, helical antennas~\cite{Lovestead19}, patch antennas~\cite{Eclercy98}, adaptive antennas~\cite{Haupt04, Laohapensaeng05} and others \cite{Haupt07}. 

GAs have previously been used in the design of various detectors and experiments, although
rarely to optimize for a science outcome directly~\cite{Liu18, Liu15}. A horn antenna was designed using a GA optimized for the detection of Cosmic Microwave Background radiation~\cite{McCarthy2016}. Both the Long-Baseline Neutrino Oscillation experiment (LBNO) and the Deep Underground Neutrino Experiment (DUNE) employed GAs to optimize the design of neutrino beamlines using simulations of a science outcome to determine the fitness~\cite{LBNO15, DUNE18}. GAs have also been used to optimize the layout of detectors, sensors, shielding, and trigger optimization~\cite{MODE, ADORE12, FLYNN10, Kleedtke, ABDULLIN2003}.

Here, we report on the initial evolution of biconical (bicone) antennas for use in radio UHE neutrino detection experiments. Using the simulated sensitivity of the antennas to neutrino interactions as the fitness score, antennas were evolved that exceed the performance of antennas now in service.  

\section{Genetic Algorithms}
A GA is an optimization technique that applies natural selection to select and generate populations of individuals so that they evolve toward an improved outcome~\cite{Holland1975, Davis1991, goldberg89, Zebulum2018, Shukla2015, Haupt1998, Beasley1993_1, FOGEL, man96, Zhang11}. Individuals are defined by their genes, which are a set of values representing the individual's characteristics. These genes form the parameter space that the GA explores. The population of individuals makes up a group of potential solutions to the problem. An individual is assessed based on its fitness score, which is the objective score that the algorithm is designed to maximize. This fitness score is the criterion used by the GA to select individuals to pass their genes to the next generation. For example, a GA may be used to optimize the volume of a box-shaped container given a constant surface area. The fitness score would then be the volume of the box, which becomes higher as the box evolves into a cube.

The next generation (a new population of individuals) is created through several stochastic and probabilistic methods, where the new individuals (children) are a variant (or a copy) of individuals from the prior generation. Selection methods decide which individuals will be used to create the children in the next generation. Genetic operators are techniques used to modify the parent's genes to generate children~\cite{Hong}. GAs are terminated when a criterion is met, such as completing a set number of generations evolved or reaching a target fitness score. 

\section{UHE Neutrinos}
One important missing piece of particle astrophysics is the detection of UHE neutrinos with energies above about $10^{17}$\,eV~\cite{Aartsen:2015rwa}. Neutrinos do not carry an electric charge and are weakly interacting, which means they can be traced back to their source more readily than other cosmic particles. However, the properties that make neutrinos resilient communicators also make them extremely difficult to detect. Their low flux of approximately tens of UHE neutrinos per \,km$^{2}$ per year per steradian~\cite{RICE, A23FourYear, AraSim, Gorham2019}, and their
interaction lengths of order $1000$\,km in the
earth~\cite{Amy_sigma,Gandhi_sigma} necessitates that experiments view on the order of $\sim100$\,km$^3$ to detect a single UHE neutrino in a year. 

Many experiments are employing antenna arrays to detect Askaryan radiation produced when a neutrino collides within a large volume of ice (such as in Antarctica or Greenland)~\cite{Askaryan, Heuge2017}. These experiments include ANITA, ARA, ARIANNA, and RNO-G, which use a variety of different antenna types~\cite{Connolly2016, Pueo21, Ara2016, Gorham2019, Anker2019, RNO}. 

ARA uses two different antenna designs to detect vertically polarized signals (VPol) and horizontally polarized (HPol) signals. ARA antennas must be designed to fit in narrow holes drilled in the ice. The ARA antennas are broadband, with the VPol antennas being birdcage bicones (13.9\,cm diameter) while the HPol antennas are ferrite-loaded, quad-slot antennas (12.7\,cm diameter)~\cite{Archambault, Allison2011, ara2019recent}. This work uses ARA as a test case, evaluating antenna performance with ARA simulation software and comparing the evolved designs to the ARA VPol antennas.

\section{The Asymmetric Bicone Antenna}
\begin{figure}[ht]
    \includegraphics[width=0.5\linewidth]{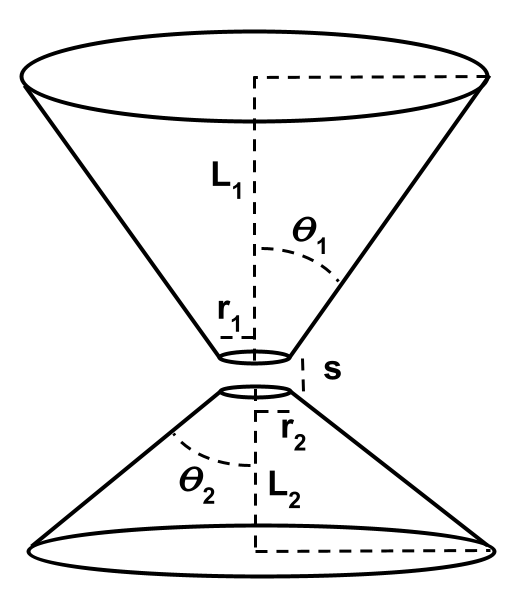}
    \caption{A schematic of an asymmetric bicone antenna. The lengths ($L_1$, $L_2$), inner radii ($r_1$, $r_2$), opening angles ($\theta_1$, $\theta_2$), and separation distance ($s$) fully define the geometry. In the results presented here, the separation distance was held constant, and the other six parameters were varied.}
    \label{fig:asymmetric_Bicone}
\end{figure}

The results presented in this paper involve the evolution of an asymmetric bicone antenna, as illustrated in Fig.~\ref{fig:asymmetric_Bicone}. A bicone antenna consists of two cones with openings facing opposite directions. This shape was chosen because it is similar to antennas currently deployed in the ARA experiment and has a broadband response, which is desirable
for the detection of the broadband 
Askaryan emission. The asymmetric bicone is fully defined by six genes (parameters): the inner radius (r), the length (L), and the opening angle ($\theta$) for the top and bottom cones. A single individual in the GA is an antenna design given by these six parameters.

The GA constrains the diameter of the antennas to 15\,cm to match the diameter of the ARA boreholes~\cite{ARADiameter}. The outer diameter of the antenna is therefore prevented from being larger than the ARA borehole width (both during initialization and in later generations). While no required borehole diameter clearance (the distance between the antenna and the borehole) was specified in the GA, ARA uses a borehole clearance of 1.1\,cm for the VPol antennas and 2.3\,cm for the HPol antennas~\cite{Archambault}. Future experiments may drill larger boreholes (over 28\,cm in diameter)~\cite{RNOGDesign}. This would improve antenna sensitivities since larger and more complex designs could be created (from the perspective of the GA, there would be a greater parameter space to explore). Here, we maintain the same borehole diameter that ARA currently uses to simplify comparisons to ARA's VPol antenna design.

The GA also constrains the minimum length of a design due to a  limitation that we have found in our simulations of the response of the antennas. We have found that the results become unreliable when a quarter of the wavelength is shorter than the length of one side of the bicone at any frequency in our band of interest.  In other words, we need to require $L > c/4f$ where $L$ is the length of the shortest cone and $c$ is the speed of light in a vacuum anywhere in the band. The lowest frequency in the bandwidth sets the shortest length that can be used at \textit{all} frequencies because it has the greatest minimum length. Higher frequencies will give shorter minimum lengths, but then those lengths will not be reliably simulated across the \textit{entire} bandwidth of interest. While 7.5 \,cm is reliable for 1000 MHz, it would be unreliable for lower $f$. The lowest frequency in the bandwidth is approximately 100\,MHz, since Askaryan signals below this frequency are dominated by galactic noise~\cite{AraSim}. For the lowest frequency in the band of interest at approximately 100\,MHz, this gives a minimum full length allowed by the algorithm of 75\,cm (each side 37.5\,cm).

\begin{figure}[h]
    \centering
    \includegraphics[width=.98\linewidth]{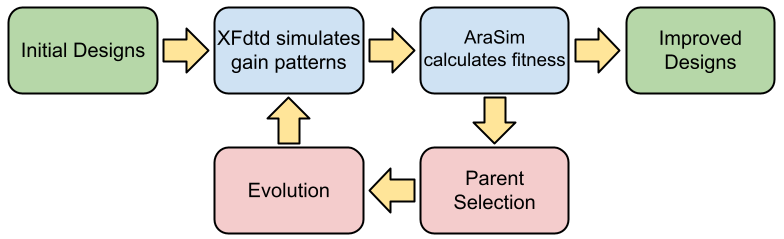}
    \caption{A diagram of the GENETIS workflow used to evolve antennas. The boxes on the far left and right give the beginning and end of the loop. The two central boxes represent the fitness calculation and the bottom two boxes represent the creation of the next generation.}
    \label{fig:GA_diagram}
\end{figure}
\section{The Genetic Algorithm}
Fig.~\ref{fig:GA_diagram} presents a schematic of the GENETIS algorithm. An initial population is generated, and this begins an iterative loop where a new generation is produced by selecting individuals from the prior generation and using variants of those individuals' genes to form the next generation. The process repeats until predetermined termination requirements are met. The following sections describe the GENETIS GA in more detail. 

\subsection{Initialization}
The first population is initialized by selecting values for the six genes for each individual from a uniform distribution with a mean similar to current ARA designs. The parameters for the initialization can be seen in Tab.~\ref{tab:intialization}. The aforementioned simulation constraint gives the minimum length. Using the minimum radius and length, the maximum angle is calculated as the angle that will cause the antenna width to equal the borehole diameter. 

\begin{table}[ht]
\caption{Range of of uniform distributions used for each gene.\label{tab:intialization}}
\begin{tabular}{l|l|l}
Gene    & Minimum    &  Maximum    \\ 
\hline \hline
Length (cm)        & 37.5 & 140 \\
Radius (cm)      & 0.0  & 7.5   \\
Opening Angle (degrees)   & 0.0 & 11.3 \\
\end{tabular}
\end{table}

\subsection{Fitness Calculation}
Once every individual in a generation is defined, the fitness score of each individual must be determined. The fitness score is a measure of performance evaluated for each individual in a generation and is used by the tournament and roulette selection methods (see Subsection C). A higher fitness score indicates that the individual performed better. The calculation of fitness scores is a multi-step process that involves two main programs integrated with the GA. First, the gain pattern of each individual is simulated. Then, a measure of the ARA detector's sensitivity to neutrino-induced radio signals is calculated by running a neutrino detection simulation software using the individual for the VPol antennas. This measure of sensitivity is the final fitness score. 

The first step in evaluating the fitness of an individual is to model its geometry in XFdtd, a commercial electromagnetic simulation software by Remcom~\cite{XF}. XFdtd simulates the antenna response at 60 different frequencies (equal steps from approximately 100\,MHz to 1000\,MHz) originating from all directions at each azimuth-zenith coordinate (in steps of $5 ^{\circ}$). An antenna's gain is a measure of how efficiently it converts received radio waves into input power. XFdtd calculates the gain of an antenna at a specific coordinate using Eq.~\ref{eq:Gain}:
\begin{equation} 
G \ = \ \frac{2\pi |E_{\theta}|^2}{\eta P_0}
\label{eq:Gain}
\end{equation}
Here, $G$ is the (absolute) gain of the antenna in a specific direction, which XFdtd reports in dBi. The electric field incident on the antenna from that direction is represented by $E$, $\eta$ is the wave impedance in the medium ($377\,\Omega$ in free space), and $P_0$ is the power accepted by the antenna.

For the second step in calculating the fitness score, a neutrino detection simulation program called AraSim is used to measure the performance of the antenna~\cite{AraSim}. Developed by the ARA collaboration, AraSim is able to model neutrinos with energies between $E_{\nu} = 10^{17}-10^{21}$\,eV~\cite{AraSim}. AraSim simulates high-energy neutrino interactions in the Antarctic ice that produce electromagnetic and hadronic showers resulting in the production of Askaryan radiation. AraSim uniformly distributes these interactions within a cylindrical volume with a 3\,km radius centered around the detector~\cite{AraSim}. The direction of the incoming neutrino is randomly distributed over a solid angle of 4$\pi$. The radio emission propagation is modeled using ray-tracing, which determines the path length from the interaction to the detector. The ray-tracing models the depth-dependent index of refraction of the ice, which is n=1.3 at the surface to n=1.8 at 200\,m deep~\cite{AraSim}. Because of this variable index of refraction, the electromagnetic waves emitted from the interaction bend en route from the interaction point to the antenna. AraSim then calculates the polarization, viewing angle, travel time at the receivers, and then models the system electronics, noise waveforms, and time-domain trigger~\cite{AraSim}.

GENETIS determines an individual's fitness score with AraSim by setting the
response of the VPol 
antennas to the individual's response generated by XFdtd for each of the 60 frequencies simulated. The sensitivity produced by AraSim, known as the effective volume, is used as the individual antenna's fitness score. The effective volume is a common quantity used to assess detector sensitivities in neutrino detection experiments making it a natural and convenient quantity. Since the effective volume is directly proportional to the number of neutrinos detected, we can directly use it as the fitness score. The effective volume $[V\Omega]_{\rm eff}$ is given by~\cite{RICE}:
\begin{equation} 
[V\Omega]_{\rm eff} \ = \ 4\pi \ V_{\rm ice} \ \frac{N_{\rm detected}}{N_{\rm simulated}}
\label{eg:veff}
\end{equation}
where $V_{\rm ice}$ is the total volume of ice simulated in AraSim, $N_{\rm detected}$ is the total number of neutrinos detected (the sum of the weights discussed above), and $N_{\rm simulated}$ is the total number of neutrinos simulated. In this analysis, $V_{\rm ice}$ is given by a cylinder around the detector with a radius of 3\,km, with a total volume of approximately 85\,km$^3$. For each individual, $N_{\rm simulated}$ is $3\times10^{5}$ neutrinos with an energy of $10^{18}$\,eV, which is in the center of ARA's region of sensitivity~\cite{AraSim}. Simulating this number of neutrinos gives a standard deviation of 0.2\,km$^3$\,sr in the
effective volume. 

The calculation of the fitness scores is a computationally heavy process and is conducted using cluster computing at the Ohio Supercomputing Center. The process is parallelized to spread $N_{\rm simulated}$ across 10 different jobs, allowing each job to be completed in approximately six hours. Because of limitations on concurrently running jobs, the fitness score calculation takes 12 hours in total per generation when evolving 50 individuals per generation and using $3 \times 10^5$ neutrinos per individual.

\subsection{New Generation Creation}

Roulette and tournament were the selection methods used in this GA. Roulette selection, also known as fitness proportionate selection, is where the probability of an individual being selected as a parent for the new generation is proportional to their fitness score~\cite{LIPOWSKI}. In tournament selection, a subset of individuals are selected, and the one with the highest fitness score is selected as a parent~\cite{Methods, Shukla2015}.

Three genetic operators were then used: reproduction, injection, and uniform crossover. Reproduction uses a selection method to obtain one parent and passes that individual directly to the next generation. The injection operator generates entirely new individuals that are not derived from any parents. Uniform crossover takes two parents from the prior generation and generates a child whose genes each have a fifty percent chance of coming from each parent~\cite{SYSWERDA}.  

The proportions for the selection methods and genetic operators were found through an optimization analysis. Different combinations of selection methods and genetic operators were tested through an exercise where an asymmetric bicone with the same six parameters as used here evolved to a predetermined geometry. The results presented here use the GENETIS algorithm with 50 individuals over 31 generations. For each new generation, 80\% of parents were selected using roulette selection and 20\% were chosen through tournament selection. Four individuals (7\% of the population) competed in each tournament. The new population was generated using 72\% crossover, 22\% injection, and 6\% reproduction. 

\subsection{Loop and Termination}
After the second generation is created, the GA continues to iterate and evolve individuals towards more optimal solutions. The loop was allowed to continue to run until it appeared that the growth in average fitness score had plateaued. Tests using the optimization analysis have shown that the majority of growth should occur by approximately the 30$^{\rm{th}}$ generation, so a plateaued mean around generation 30 indicates that there is little or no growth remaining. 

\section{Results}
\subsection{Results From Loop}
\begin{figure}[!ht]
    \centering
    \includegraphics[width=1\linewidth]{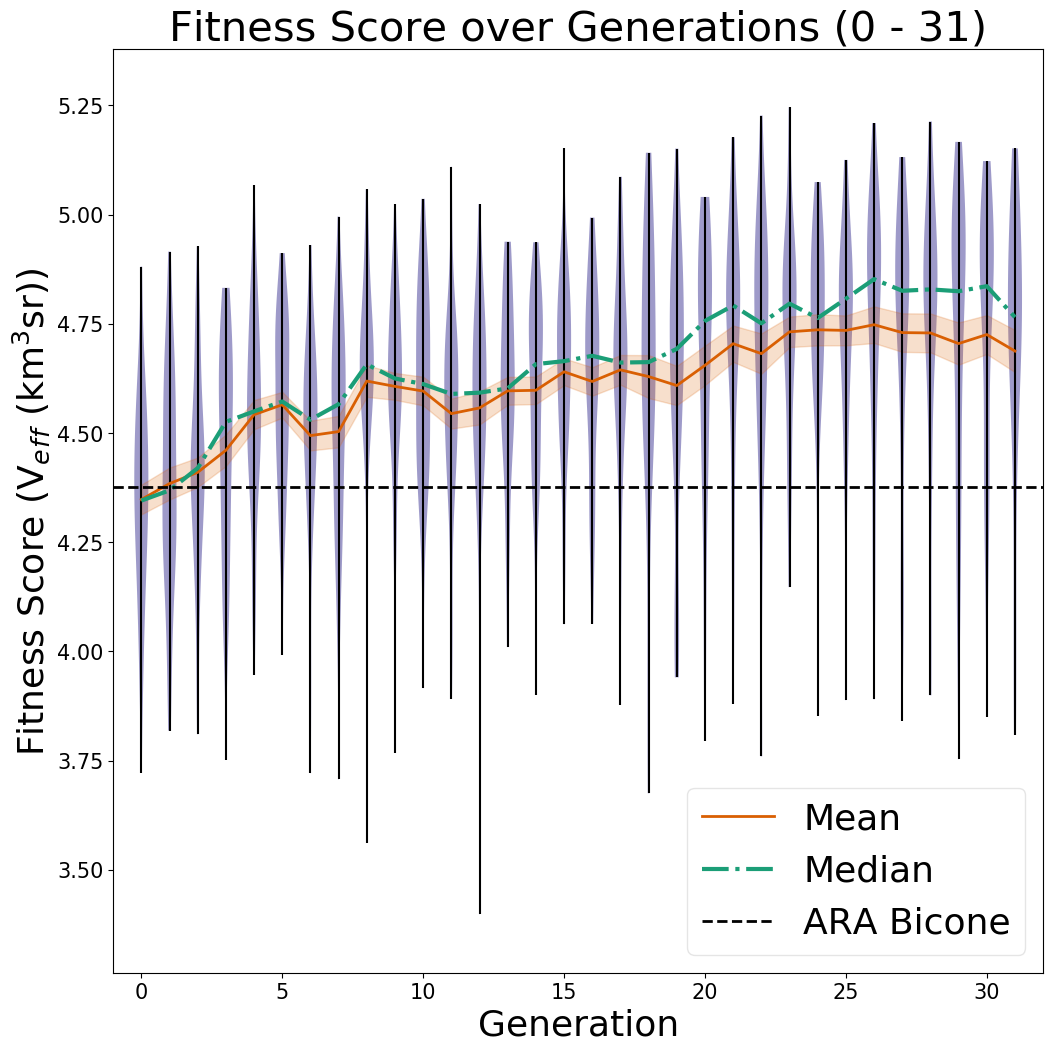}
    \caption{Initial results of GA. Each violin represents the entire range of scores in the generation, while width indicates the density of scores. The current ARA Bicone Fitness is shown as the horizontal dotted line.}
    \label{fig:violin}
\end{figure}

\begin{figure}[ht]
    \centering
    \includegraphics[width=1\linewidth]{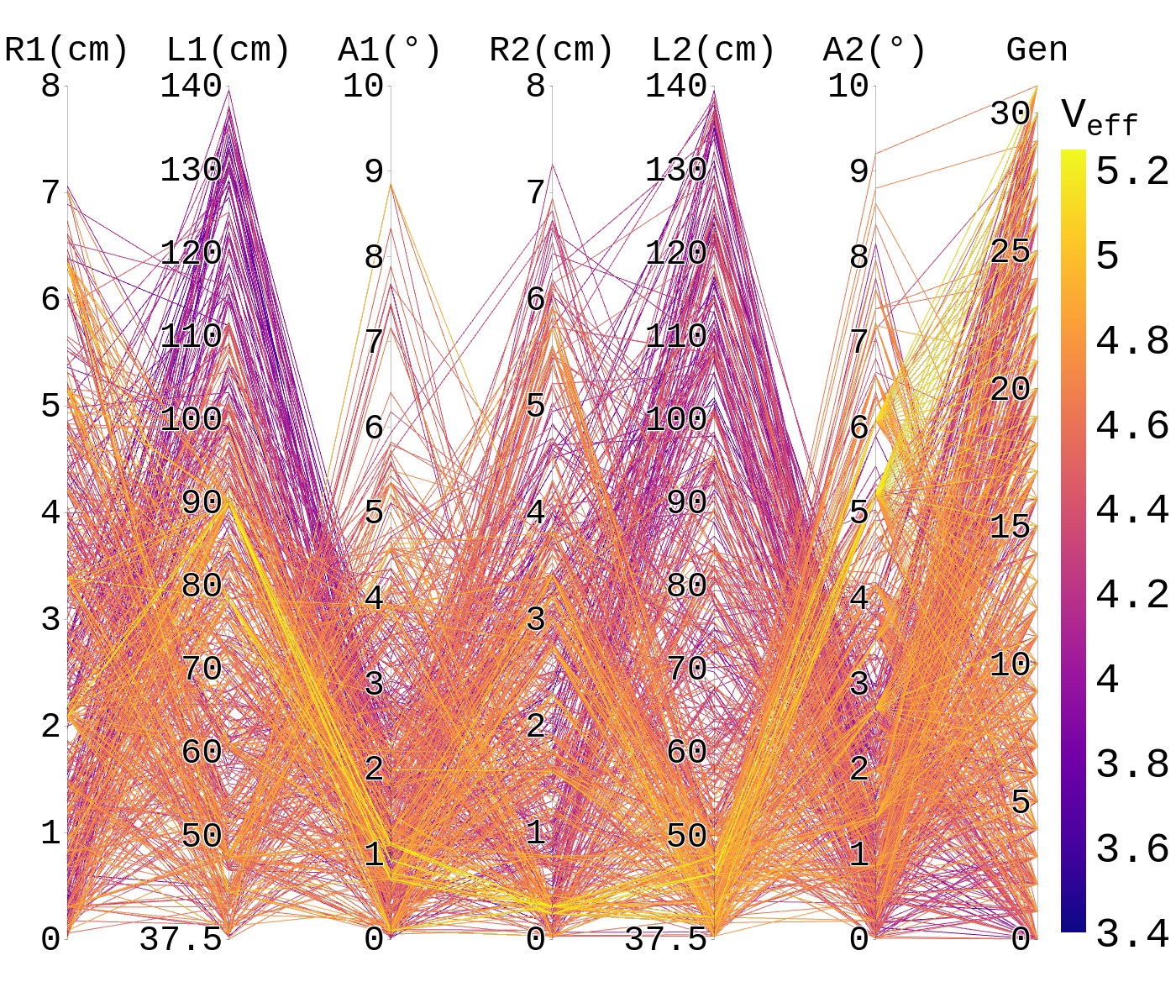}
    \caption{ Evolution of the six antenna parameters optimized so far, showing trends toward preferred features (bright yellow being most fit).  }
    \label{fig:rainbow}
\end{figure}

The results of the evolution are presented in the violin plot in Fig.~\ref{fig:violin} showing clear evolution toward improved solutions. The top point of each generation shows the highest fitness score of that generation. The overall highest scoring antenna occurred in generation 23 with a fitness score of $5.2\,\pm\,0.2$\,km$^3$\,sr, which is 22\% higher than the score when ARA's current VPol design is used. For each generation, the range of fitness scores is illustrated by the height of the violin. The width of each violin represents the density of individuals with that score. The solid orange line shows the mean of the population, with the standard deviation on the mean represented by the orange shading, and the dashed green line shows the median, which is useful for understanding the convergence of the population. Lower scoring individuals are still present throughout the entire evolution, despite the average and maximum fitness score improving beyond the initial generation. This is primarily due to the injection operator, which continually introduces new diversity to the population to prevent early convergence to local maxima~\cite{immigration}. 
Fig.~\ref{fig:rainbow} is a parallel coordinate plot that shows the evolution of each gene over the entire run. An individual is represented by a jagged line spanning the width of the plot, with the value of each of the individual's genes represented by the line's height on the vertical axes. The color of the line represents the individual's fitness score. This demonstrates the effectiveness of the GA at producing higher scoring individuals in later generations. 

Fig.~\ref{fig:rainbow} also shows general trends in each of the genes and their impact on the fitness score. For example, most high scoring antennas share similar values, with opening angles of the top cone (A1) being under 1 degree, the length of the bottom cone (L2) being less than 50\,cm, and the angle of the bottom cone (A2) being between 4 and 6 degrees for high scoring antennas. However, the other parameters have a larger spread in viable values, with the radius of the top cone (R1) being spread across the entire parameter space for high-scoring individuals. The fitness score was not expected to depend highly on the inner radii due to the small range of sizes available for the radii.

\begin{figure}[!h]
    \centering
    \includegraphics[width=.2\linewidth]{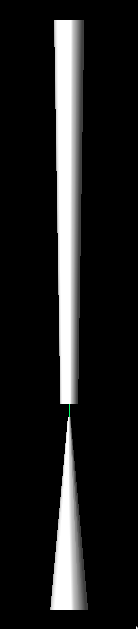}
    \caption{Model of the best antenna design. Individual 8, evolved in Generation 23.  Other individuals are not shown because they are not visually distinguishable
    from this one.}
    \label{fig:antennas}
\end{figure}

Fig.~\ref{fig:antennas} shows a 3D model of the highest performing antenna evolved in this work, and its genes are represented in Tab.\ref{tab:TopIndividual}. Notice that the top section of the antenna is longer than the bottom, has a larger inner radius, and has a smaller opening angle. This is also true of the next best antennas, which had similar genes to the highest performing individual. Of the five highest scoring antennas, only a combined three genes (out of thirty) were more than 5\% different from the highest scoring individual. 

\begin{table}[!ht]
\caption{Genes of the best individual in the evolution.\label{tab:TopIndividual}}
\begin{tabular}{l|l|l|l}
Section & Radius (\,cm)  & Length (\,cm)   &  Angle ($^{\circ}$)    \\ 
\hline \hline
Top  & 2.09 & 89.9 & 0.927 \\
Bottom & 0.302 & 45.4 & 5.22   \\
\end{tabular}
\end{table}

\subsection{Comparing Gain and Realized Gain}
As discussed in Section V.B., the antenna response is calculated in XFdtd using Eq.~\ref{eq:Gain}. While the antenna response is not the fitness score, AraSim must use it to evaluate the fitness score. In addition to calculating the antenna gain, XFdtd also calculates the \textit{realized gain}. Realized gain accounts for the reflection of a received signal due to an impedance mismatch~\cite{realizedgain}. The reflection coefficient affects the power that reaches the antenna, $P_0$. Given a power $P_{\rm M}$ reaching the matched transmission line, we have~\cite{IEEE1979}:

\begin{equation} 
P_0 \  =  P_{\rm M} (1 - \Gamma^2 ) 
\label{DeliveredPower}
\end{equation}

\noindent where $\Gamma$ is the reflection coefficient. In the case where there is no impedance mismatch, the realized gain is equivalent to the gain. The realized gain $G_R$ in XFdtd is given by the following equation, which replaces $P_0$ from Eq.~\ref{eq:Gain} with $P_{\rm M}$:

\begin{equation} 
G_{\rm R} \  =  \ \frac{2\pi |E_{\theta}|^2}{\eta P_{\rm M}} \ .
\label{RealizedGain}
\end{equation}

\begin{figure}[!ht]
    \centering
    \includegraphics[width=\linewidth]{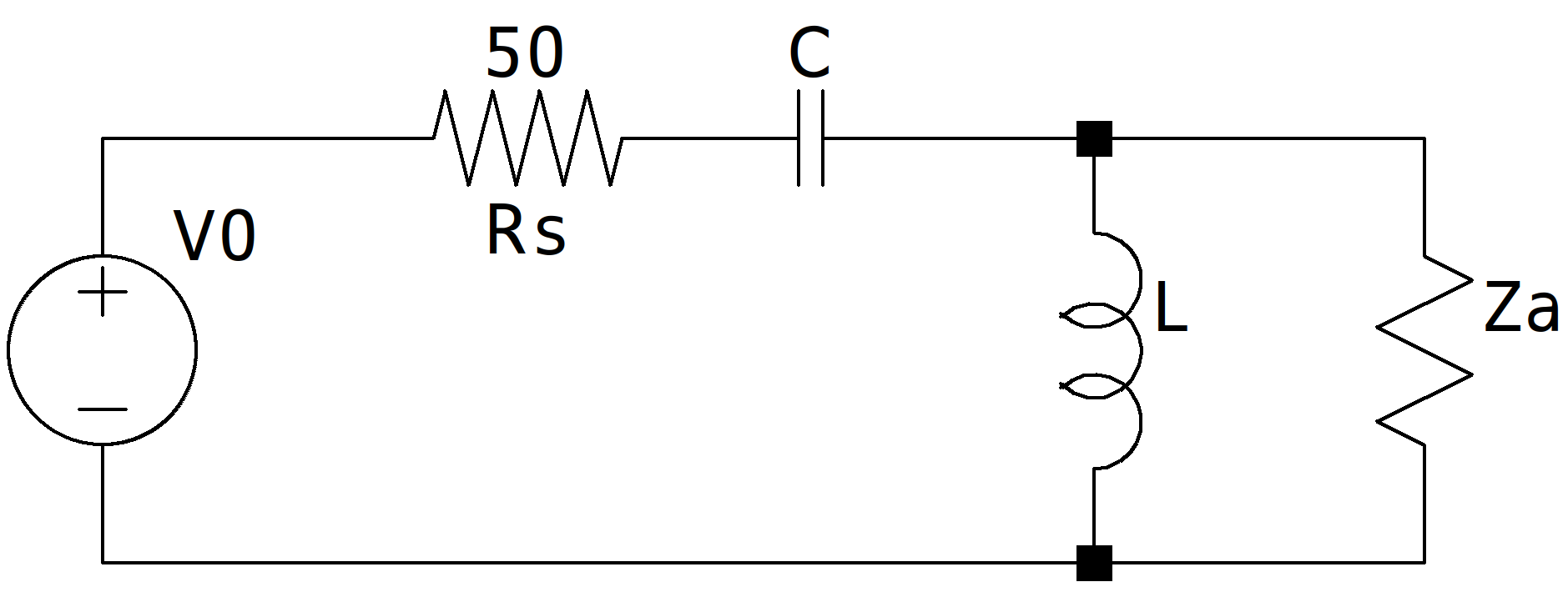}
    \caption{SPICE schematic of the matching circuit.}
    \label{fig:SPICE}
\end{figure}

Minimal reflection occurs when the load and characteristic impedances are equal, as $\Gamma = \frac{Z_a-Z_0}{Z_a+Z0}$ for a load $Z_a$ and a source $Z_0$. For this work, a simple matching circuit was designed to match the impedance of the load to the source at a single frequency (200\,MHz), as shown in the SPICE schematic in Fig.~\ref{fig:SPICE} ~\cite{spice}. Given a load impedance of $Z_a = R_a + i X_a$, a source impedance of $Z_s = R_s$, and an angular frequency of $\omega = 2\pi f$, the inductance and capacitance of the matching circuit components can be derived (see Appendix~\ref{sec:derivation}) to be~\cite{Niknejad_2007}: 

\begin{equation} 
\mathcal{L} \ = \ \frac{\sqrt{R_a(R_s R_a^2 - R_s^2 R_a + X_a^2 R_s)} + X_a R_s}{\omega (R_a-R_s)} \\
\label{eq:Inductance}
\end{equation}
\begin{equation} 
C \ = \ \sqrt{\frac{R_a}{\omega^2(R_s R_a^2 - R_s^2 R_a + X_a^2 R_s)}} \ .
\label{eq:Capacitance}
\end{equation}

\begin{figure}[!ht]
    \centering
    \includegraphics[width=\linewidth]{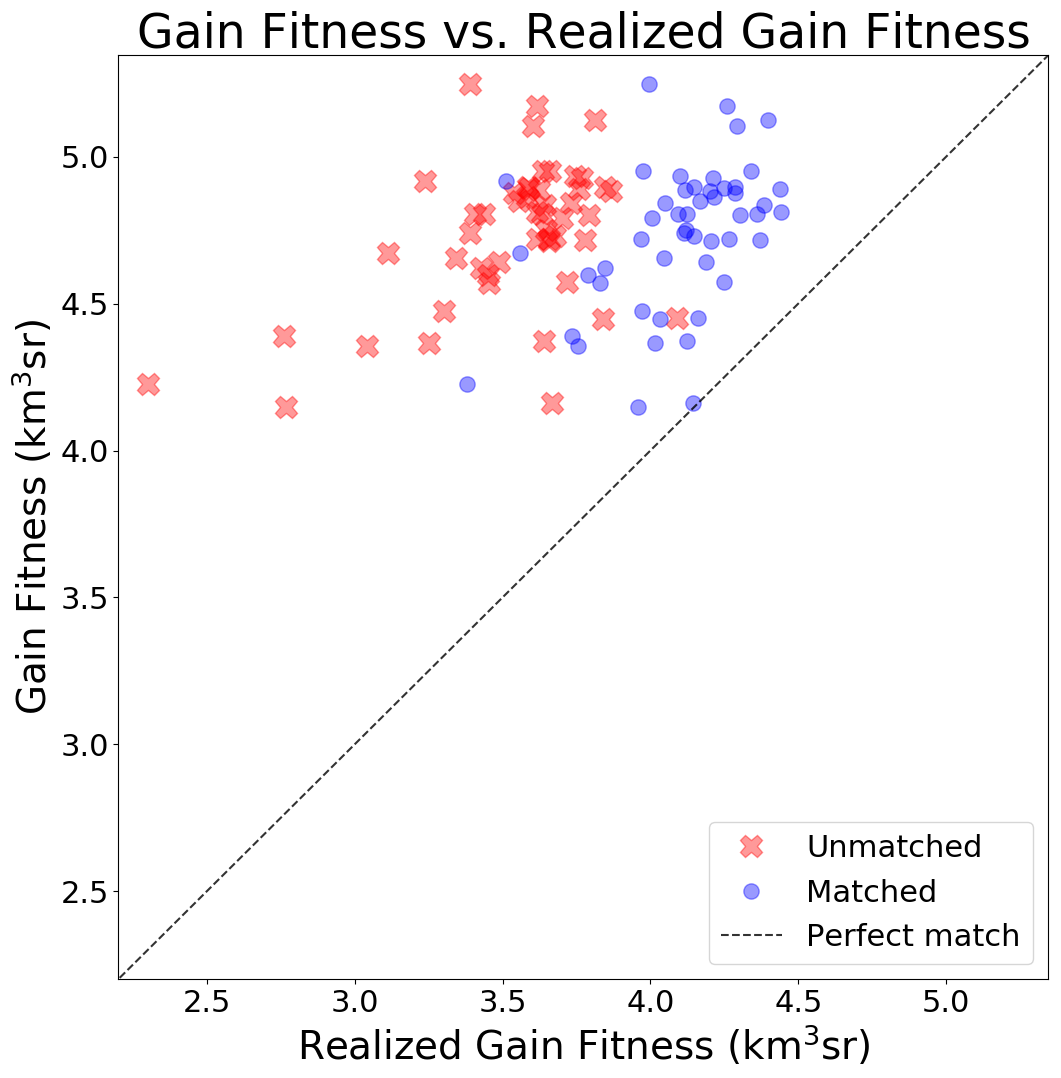}
    \caption{Fitness scores using gain vs. realized gain with a matching circuit (blue circles) and without (red crosses).}
    \label{fig:Gain_vs_Realized_unmatched}
\end{figure}

\begin{figure}[!ht]
    \centering
    \includegraphics[width=\linewidth]{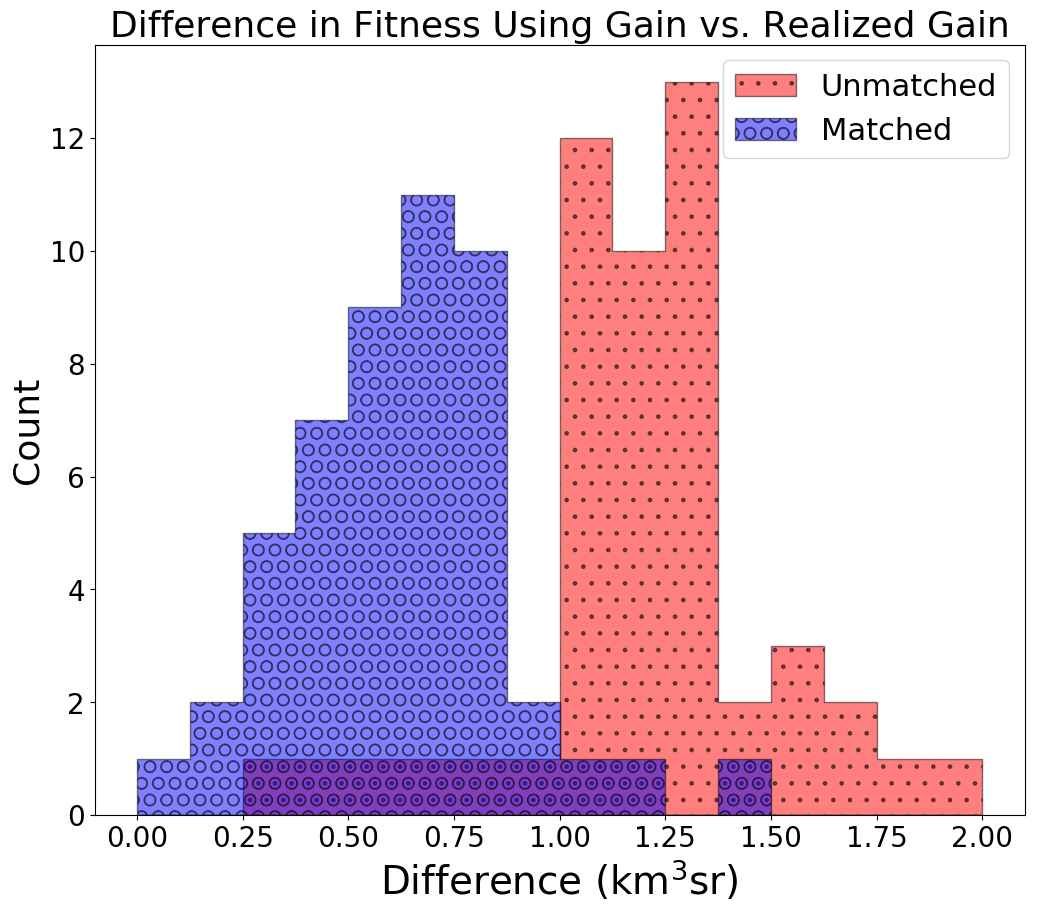}
    \caption{Difference in fitness scores using gain vs. realized gain (red/dots) and matched realized gain (blue/circles).}
    \label{fig:Fitness_Difference_unmatched}
\end{figure}

\begin{figure*}[t]
    \centering
    \includegraphics[width=.47\textwidth]{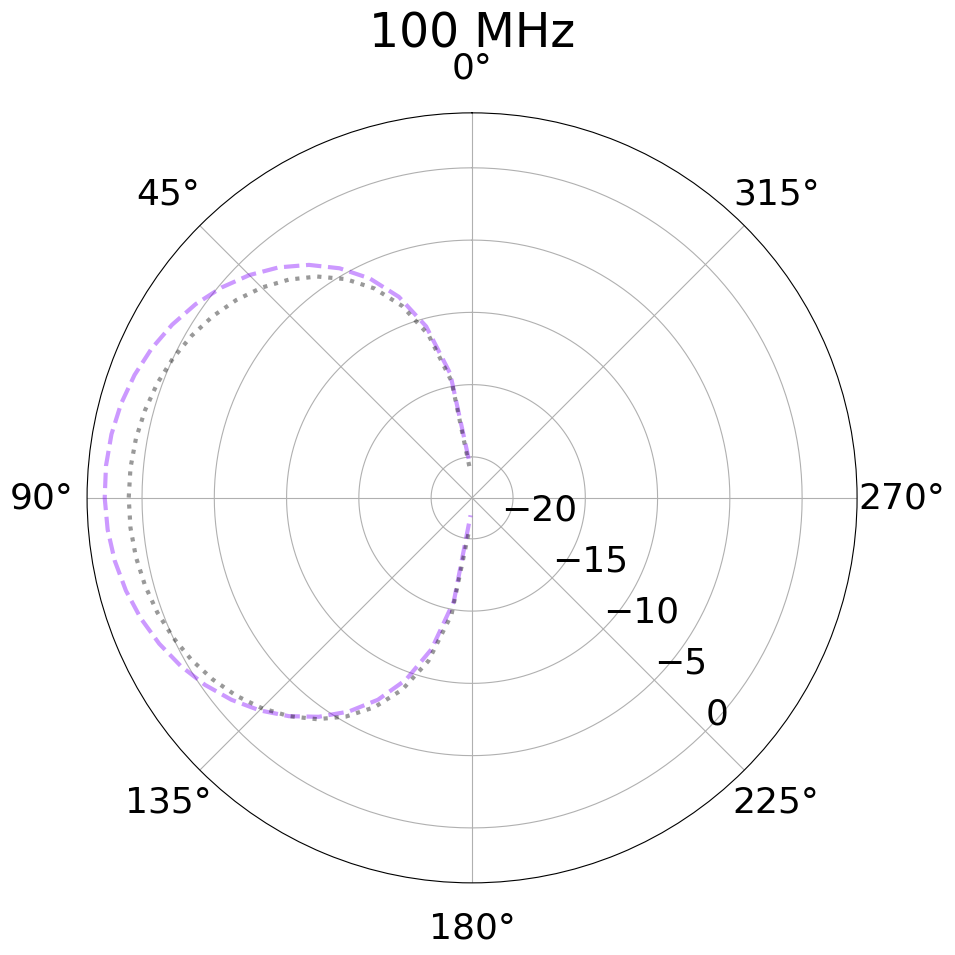}\hspace{.2 in}
    \includegraphics[width=.47\textwidth]{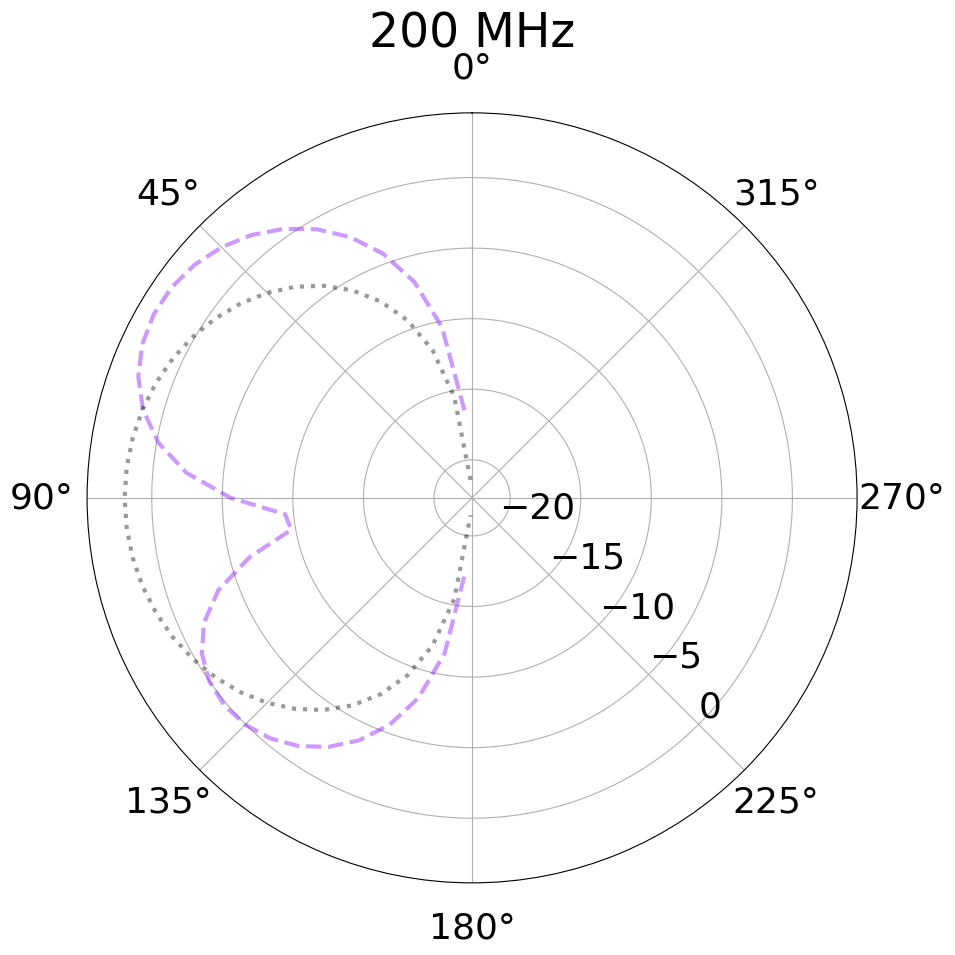}
    \\[\smallskipamount]
    \includegraphics[width=.47\textwidth]{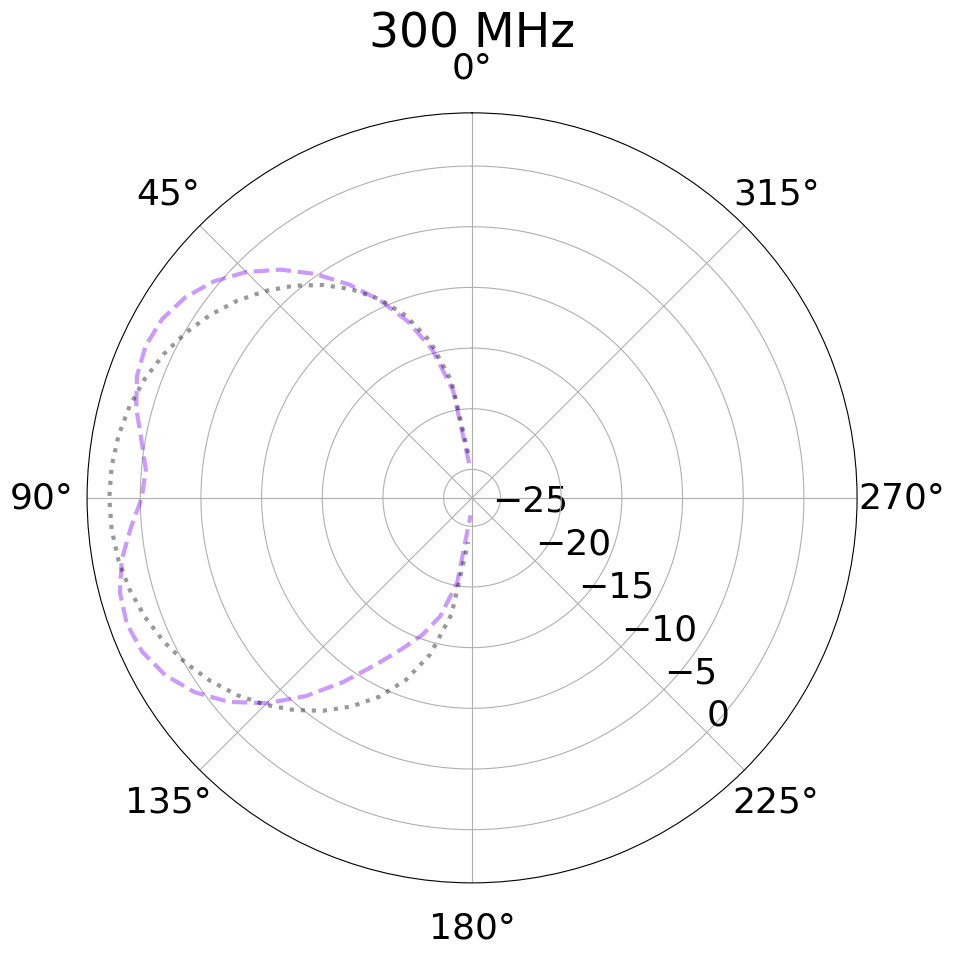}\hspace{.2 in}
    \includegraphics[width=.47\textwidth]{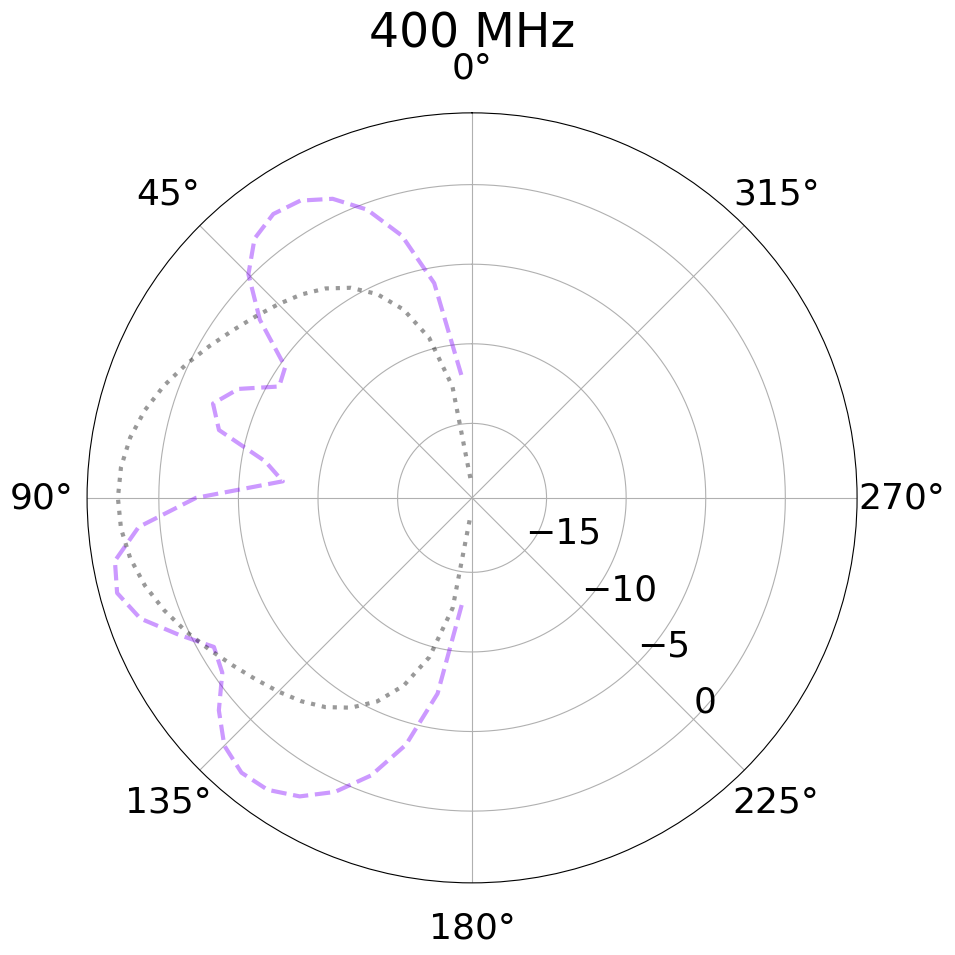}
    \\[\smallskipamount]
        \caption{Antenna response patterns for the evolved bicone (purple dashed line) and the ARA bicone (black dotted line) from XFdtd. Angles are measured from the positive vertical direction.}
    \label{fig:ARAGainCompare}
\end{figure*}

Fig.~\ref{fig:Gain_vs_Realized_unmatched} compares the fitness scores of the individuals from generation 23 when evaluated using the absolute gain (hereafter referred to as simply ``gain") to the fitness scores when evaluated with the realized gain. For the realized gain, we use a custom matching circuit for each individual designed as described above for 200\,MHz.
Naturally, the performance decreases when using realized gain; however, the matching circuit mitigates this effect.

Fig.~\ref{fig:Fitness_Difference_unmatched} shows a histogram of the differences in fitness scores for each individual in generation 23 when calculated with gain and realized gain for both the unmatched and matched cases. The unmatched and matched scores are, on average, 1.2\,km$^3$\,sr and 0.6\,km$^3$\,sr lower than the scores evaluated with absolute gain. The highest performing individual in generation 23 falls from a fitness score of 5.25 to 4.00\,km$^3$\,sr when evaluated with realized gain using the matching circuit. Four antennas still exceeded the ARA bicone's score of 4.38\,km$^3$\,sr when applying the matching circuit, which was only evaluated using gain.

\subsection{Physics Interpretation of Results}

\newlength{\imagewidthtwo}
\newcommand{\subgraphicstwo}[2]{
\settowidth{\imagewidthtwo}{\includegraphics[height=8cm]{#1}}%
\begin{subfigure}{\imagewidthtwo}%
    \includegraphics[height=8cm]{#1}%
    \caption*{#2}%
\end{subfigure}%
}

\begin{figure*}[!ht]
    \centering
        \subgraphicstwo{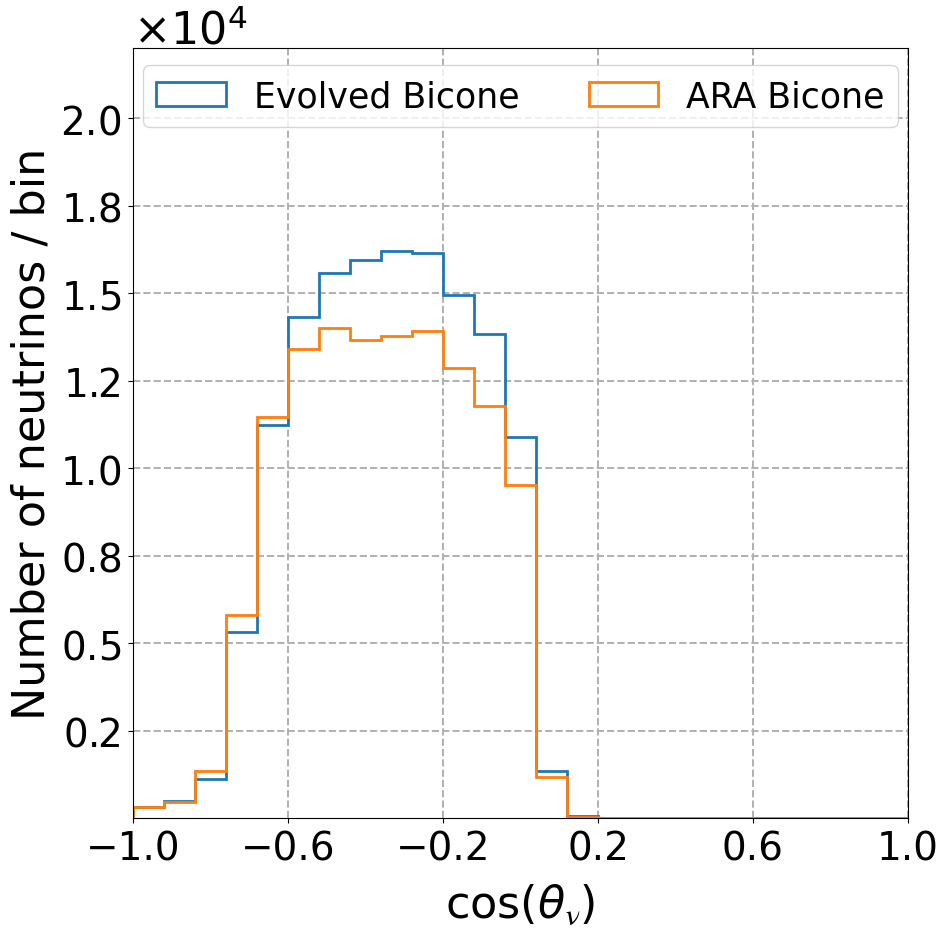}{}
        \subgraphicstwo{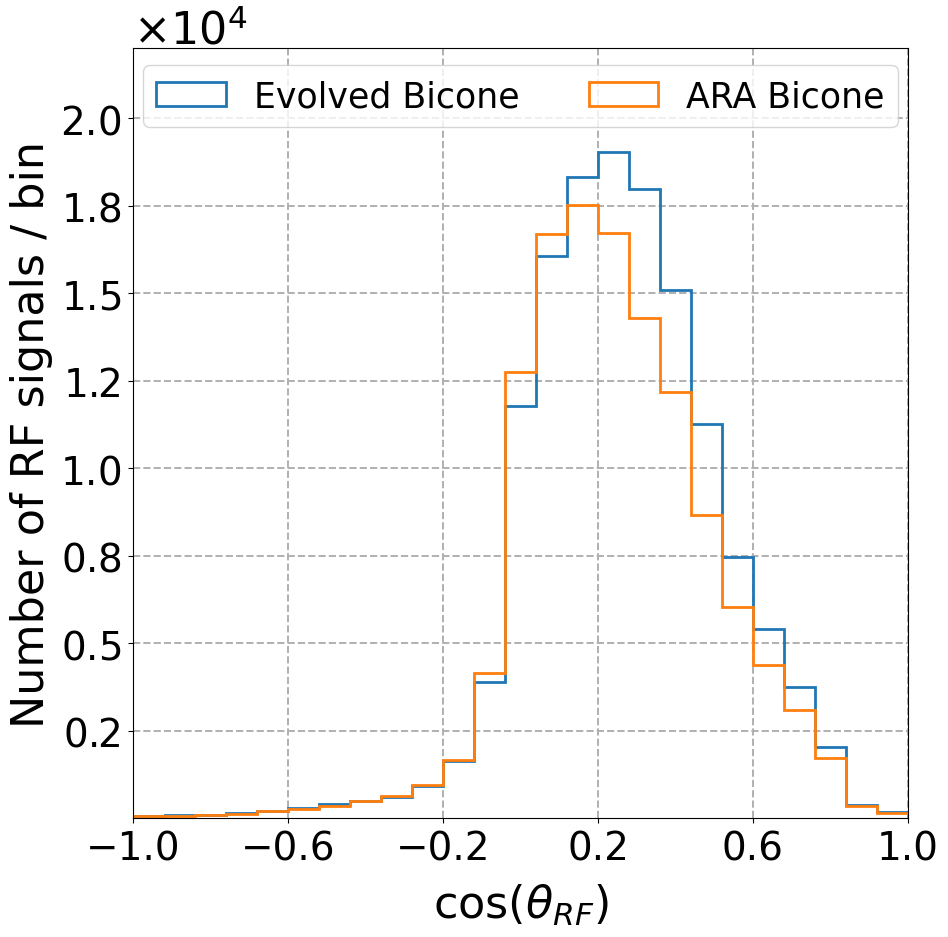}{}
        \caption{Histograms of the number of detected neutrinos (left) and RF signals (right) by each bicone for $3\times10^7$ simulated events.  For
        $\cos{\theta_{\nu}} = 1$ ($\theta_{\nu}=0^{\circ}$), the neutrino originated from below the detector (up-going), and for
        $\cos{\theta_{\rm RF}}=1$
($\theta_{\rm RF}=0^{\circ}$),  the radio
signal is incident from below.}
    \label{fig:DirectionCompare}
\end{figure*}

It is important to determine what causes the improvement observed in fitness scores when using absolute gain. One notable difference between the 
antenna evolved here and the one used by ARA 
is the asymmetry of the geometry, which results in a  qualitative difference in the shape of the antenna responses.

Fig.~\ref{fig:ARAGainCompare} shows the gain patterns of the best performing individual in the evolution, individual 8 in generation 23, compared to the gain pattern of the ARA V-pol antenna at four frequencies.
The beam patterns show modestly higher peak
gains (1-2\,dB at low frequencies to a few\,dB
at higher frequencies). While at 200\,MHz the peak gain is actually at about 50$^{\circ}$ from the vertical, a preference for signals from below the surface becomes evident at higher frequencies, as can be seen in the 400\,MHz gain pattern.

Fig.~\ref{fig:DirectionCompare} shows two histograms comparing the evolved bicone and the ARA bicone for a simulation with $3 \times 10^7$ neutrinos at $10^{18}$\,eV. The left panel of Fig.~\ref{fig:DirectionCompare} presents the number of detected neutrinos by ARA when each of the two antennas designed are used, binned by the cosine of the zenith angle of the neutrino's trajectory, where $\cos{\theta_{\nu}} = 1$ ($\theta_{\nu}=0$) indicates that the neutrino originated from below the detector (up-going). Few events are detected by ARA with a cosine angle greater than about 0.2 ($\theta_{\nu}$ less than about $80^{\circ}$) due to the absorption of those events in the Earth. When the bicone is used, AraSim predicted 14\% more detected events with neutrinos incident at angles between the horizontal and $37^{\circ}$ below horizontal than with the current ARA design.

\begin{figure}[!ht]
    \centering
    \includegraphics[width=1\linewidth]{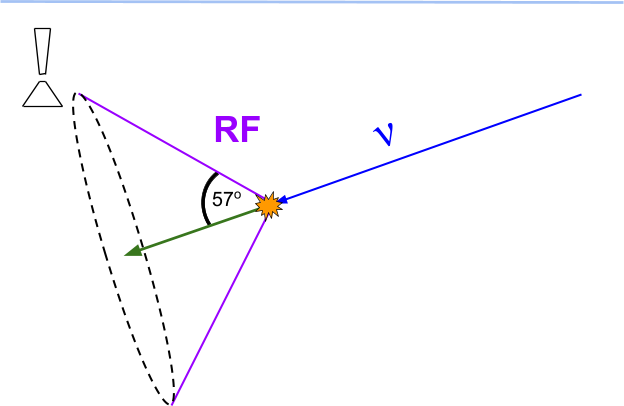}
    \caption{A down-going neutrino interacting in the ice below the antenna, creating a particle cascade that produces Askaryan radiation that is viewed by the antenna.}
    \label{fig:Downgoing_Nu}
\end{figure}

The right panel of Fig.~\ref{fig:DirectionCompare} also shows the number of detected
neutrinos, now binned by the cosine of the zenith angle of the direction of the RF signal as it is received. Here, $\cos{\theta_{\rm RF}}=1$($\theta_{\rm RF}=0^{\circ}$) indicates that the radio signal is incident from below.
ARA with the evolved bicone is predicted to detect more RF signals between cosine angles of 0.2 and 0.6 (about 50-80$^{\circ}$ from the vertical), meaning more signals were detected at angles originating from events below horizontal. These results are consistent with an improvement in the detection of down-going neutrinos that interact in the ice and produce radio signals that propagate up to the detectors. This is shown in Fig.~\ref{fig:Downgoing_Nu}, where a down-going neutrino interacts in the ice and creates a particle cascade. The resulting lepton moves through the ice faster than the speed of light (in ice), creating a cone of Askaryan radiation.

\section{Conclusion}
With these results, GENETIS presents the results of a GA-designed antenna using a physics outcome as a measure of fitness and lays the foundation for future detector optimizations. We show that a GA evolving six parameters of a bicone antenna can evolve a design that results in an ARA detector with a 22\% greater sensitivity to in-ice UHE neutrino detection than one using the current VPol antennas. The improved design outperforms the existing ARA VPol antenna in detecting down-going neutrinos that produce radio waves that propagate up towards the detector.

\section{Future Work}
An antenna prototype of the best performing individual from this work will be fabricated through additive manufacturing at The Ohio State University Center for Design and Manufacturing Excellence. This will allow us to compare laboratory measurements to the results of the simulations produced by GENETIS. In future work, we will continue to evolve improved antennas with the goal of doubling the current ARA VPol antenna sensitivity. If an evolved antenna successfully improves on the current VPol antennas by a factor of two, it will be deployed in-ice for further testing.

The GENETIS collaboration is currently working on several improvements to the GA to further improve the computational efficiency, convergence speed, and maximum fitness. First, we are introducing more complex antenna geometries, such as bicones with nonlinear sides. These antennas would require additional genes that describe the coefficients of polynomials that represent the shape of the sides of the bicone. This project is underway, and the evolution of other types of antennas is also in development.  We are also exploring the use of additional and more advanced selection methods and genetic operations, including rank selection and elitism.

Additionally, we will continue to refine our workflow to improve the sophistication of our modeling, such as by constructing broadband matching circuits for evolved antennas. The construction of a single frequency matching circuit can currently be automated in the loop, allowing us to evolve using realized gain directly. We plan to fully implement this with broadband matching circuits to more realistically and efficiently evolve designs.  

In the future, the GENETIS project will expand beyond antenna design and explore other aspects of experimental design and analysis,
including detector layouts and trigger
optimization. 
As a first step toward this goal, we will develop the capability to evolve the layout of an 
array of antenna stations for UHE detection together with the antenna designs.
The GENETIS project will also expand to employ different types of computational intelligence and machine learning techniques and  perform optimizations for other experimental applications.  

The successful deployment of GA-designed detectors could pave the way for additional applications of optimization heuristics for the design of scientific instruments. Expanded research in this area will streamline the optimization of the design of many types of experiments across fields for superior science outcomes.

\begin{acknowledgments}
The GENETIS team is grateful for support from the Ohio State Department of Physics Summer Undergraduate Research Program, support from the Center for Cosmology and Astroparticle Physics, and the Cal Poly Connect Grant. We would also like to thank the Ohio Supercomputing Center. J. Rolla would like to thank the National Science Foundation for support under Award 1806923 and the Ohio State University Alumni Grants for Graduate Research and Scholarship. We are grateful to the ARA Collaboration for making available the AraSim simulation program used in this work, and for helpful feedback from the collaboration. 
GENETIS is grateful to Prof. Chi-Chih Chen of the
Ohio State University for his feedback on the project and paper
drafts.  Additionally, we thank 
 Dr. Brian Clark
of Michigan State University, Dr. Jorge Torres
of Yale University, and Dr. Steven Prohira 
of the Ohio State University for the valuable help
that they have given to GENETIS.  We acknowledge
Dr. Edward Herderick of the Ohio State University and the Center for Design and Manufacturing Excellence (CDME) for his thoughtful input to the project
and Prof. Dean Arakaki from Cal Polytechnic State University 
for his contributions to GENETIS.
We also acknowledge and are grateful to
the following students who made contributions
to GENETIS in its earlier phases:  Adam Blenk,  Max Clowdus, Suren Gourapura, Corey Harris, Hannah Hassan, Parker Kuzma, Luke Letwin, David Liu, and Jordan Potter, Cade Sbrocco, and Jacob Trevithick. 

\end{acknowledgments}

\bibliography{main}

\clearpage

\import{./}{Appendix.tex}

\end{document}

%% file: GENETIS_institutes.tex
\newcommand{\atOSU}{\affiliation{Dept. of Physics, Center for Cosmology and AstroParticle Physics, The Ohio State University, Columbus, OH 43210}}
\newcommand{\atUA}{\affiliation{University of Arizona, Biosphere 2, S Biosphere Rd, Oracle, AZ 85623}}
\newcommand{\atDenison}{\affiliation{Dept. of Physics and Astronomy, Denison University, Granville, OH 43023}}
\newcommand{\atCalPoly}{\affiliation{Physics Dept., California Polytechnic State University, San Luis Obispo, CA 93407}}
\newcommand{\atPSU}{\affiliation{Dept. of Physics, Dept. of Astronomy and Astrophysics, Pennsylvania State University, State College, PA 16802}}
\newcommand{\atMSU}{\affiliation{Dept. of Computer Science and Engineering, Michigan State University, East Lansing, MI 48824}}
\newcommand{\atJPL}{\affiliation{Jet Propulsion Laboratory, NASA, Pasadena, CA 91109}}

%% file: GENETIS_authors.tex
\author{J.~Rolla} \email[Email: ] {julie.a.rolla@jpl.nasa.gov} 
\altaffiliation[Work performed at The Ohio State University, currently at ]
{JPL.}
\atOSU \atJPL

\author{A.~Machtay}\atOSU
\author{A~Patton}\atOSU

\author{W.~Banzhaf}\atMSU
\author{A.~Connolly} \email[Email: ] {connolly@physics.osu.edu} \atOSU

\author{R.~Debolt}\atOSU
\author{L.~Deer}\atOSU
\author{E.~Fahimi}\atOSU
\author{E.~Ferstle}\atOSU
\author{P.~Kuzma}\atOSU
\author{C.~Pfendner}\atDenison
\author{B.~Sipe}\atOSU
\author{K.~Staats}\atUA
\author{S.A.~Wissel}\atCalPoly\atPSU

\collaboration{The GENETIS Collaboration}\noaffiliation

%% file: Appendix.tex
\appendix
\section{Impedance Matching}
\label{sec:derivation}

The following is a derivation of Eq.~\ref{eq:Inductance} and \ref{eq:Capacitance} for the elements of a single frequency matching circuit. We begin with a source resistor with impedance $Z_s = R_s$ and a load (antenna) impedance $Z_a = R_a + iX_a$. 

In summary, we seek to use purely reactive circuit components to minimize reflection and deliver all of the power from the source to the load. To do this, we will first construct a parallel sub-circuit (the parallel inductor-load resistor in Fig.~\ref{fig:SPICE}) which has a resistance equal to the resistance of the source resistor. In this case, the parallel component will be an inductor.

This parallel subcircuit has impedance 
\begin{equation}
Z_p \ = \ R_p + iX_p \ = \ (\frac{1}{R_a + iX_a} + \frac{1}{i \omega \mathcal{L}})^{-1}
\label{eq:Subcircuit}
\end{equation}

\noindent where $i \omega L$ is the impedance of the inductor. We want to find an inductance such that $R_p = R_s$. We can rearrange Eq.~\ref{eq:Subcircuit} to solve for the impedance of the inductor:

\begin{equation}
i \omega \mathcal{L} \ = \ \frac{R_pR_a - X_pX_a + i(X_pR_a + R_pX_a)}{R_a - R_p + i(X_a - X_p)} \ .
\label{eq:Ind_Imp}
\end{equation}

\noindent Eq.~\ref{eq:Ind_Imp} gives the unknown $L$ in terms of another unknown $X_p$. To simplify, we can rewrite it as

\begin{equation}
i \omega \mathcal{L} \ = \ \frac{A + iB}{C + iD}
\label{eq:Ind_Imp_Sub}
\end{equation}

\noindent using the substitutions

\begin{gather*}
A \ = \ R_pR_a - X_pX_a \\
B \ = \ X_pR_a + R_pX_a \\
C \ = \ R_a - R_p \\
D \ = \ X_a - X_p \ .
\end{gather*}

\noindent We can further rewrite Eq.~\ref{eq:Ind_Imp_Sub} so that the denominator is purely real:

\begin{equation}
i \omega \mathcal{L} \ = \ \frac{AC + BD + i(BC - AD)}{C^2 + D^2}.
\label{eq:Ind_Imp_Sub_Re}
\end{equation}

Since the impedance of the inductor must be purely reactive (imaginary), we obtain a second equation to constrain our two unknowns: $AC+BD = 0$. We substitute $D = -AC/B$ into Eq.~\ref{eq:Ind_Imp_Sub_Re} to obtain

\begin{equation}
i \omega \mathcal{L} \ = \ \frac{iB}{C} \ .
\label{eq:Ind_Imp_Final}
\end{equation}

\noindent Using the condition $D = -\frac{AC}{B}$, we can solve for $X_p$:

\begin{equation}
X_p \ = \ \sqrt{\frac{R_p R_a^2 + X_a^2 R_p - R_a R_p^2}{R_a}} \ . 
\label{eq:ParallelReactance}
\end{equation} \\

\noindent Finally, using Eq.~\ref{eq:ParallelReactance} in $B$, substituting $B$ and $C$ back into Eq.~\ref{eq:Ind_Imp_Final}, setting $R_p = R_s$, and solving for $L$ yields Eq.~\ref{eq:Inductance}.

Using an inductor with inductance given by Eq.~\ref{eq:Inductance} gives a parallel subcircuit with $Z_p = R_p + iX_p = R_s + iX_p$. Now, we will use a capacitor to offset the reactive component of $Z_p$ so that $Z = Z_p + Z_c = R_s$. This gives us

\begin{equation}
Z = Z_p + Z_c = R_s + iX_p + R_c + iX_c \ .
\label{eq:FullImpedance}
\end{equation}

Since Z should be purely real, the reactance of the capacitor is given by

\begin{equation}
X_c = -X_p\ .
\label{eq:CapacitorReactance}
\end{equation}

\noindent The reactance of a capacitor is $X_c = \frac{1}{\omega C}$, so Eq.~\ref{eq:CapacitorReactance} gives us

\begin{equation}
C = \frac{1}{\omega X_p}.
\label{eq:Capacitance2}
\end{equation}

\noindent Note that the capacitance is negative here. The negative sign indicates that the capacitor serves to \textit{decrease} the circuit's reactance. Substituting $X_p$ from Eq.~\ref{eq:ParallelReactance} into Eq.~\ref{eq:Capacitance2} yields Eq.~\ref{eq:Capacitance}.

 \begin{figure}[!ht]
    \centering
    \includegraphics[width=0.85\linewidth]{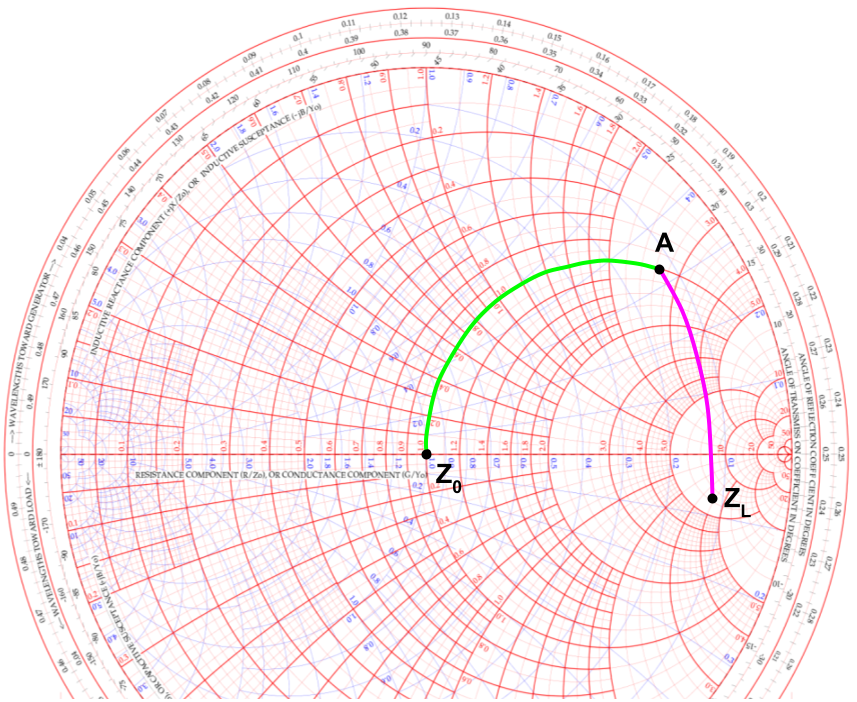}
    \caption{Smith chart example matching load $Z_L$ to source $Z_0$~\cite{Julie}}
    \label{fig:SmithChart}
\end{figure}

This process of impedance matching can be visualized using a Smith chart~\cite{Smith_Chart}. Fig.~\ref{fig:SmithChart}~\cite{Julie}  is a Smith chart showing a load impedance $Z_L$ and a source impedance $Z_0$. The purple path connecting $Z_L$ to $A$ represents the shunt inductor we designed. The inductor in parallel with the load forms a subcircuit with a resistance equal to the resistance of the source, represented by the red circle passing through $Z_0$ and $A$. The series capacitor, represented by the green path, moves us along the red circle by changing the circuit's reactance until it matches the reactance of $Z_0$.